# Efficient Picosecond-Laser Lift-Off of Copper Oxide from Copper: Modelling and Experiment

Andrius Žemaitis, Paulius Gečys, Mindaugas Gedvilas*

*Department of Laser Technologies (LTS), Center for Physical Sciences and Technology (FTMC), Savanorių Ave. 231, 02300 Vilnius, Lithuania*

*Corresponding author: mindaugas.gedvilas@ftmc.lt*

**Abstract**

Laser-induced lift-off of functional surface layers is a key process in micro- and nano-fabrication; however, optimization criteria for maximizing the lifted-off area remain insufficiently defined. In analogy to the well-established theory of efficient laser ablation, where the maximum ablated volume per pulse is achieved at a peak fluence of $F_0^{\text{opt}} = e^2 F_{\text{th}}$, we develop a theoretical framework for efficient laser lift-off driven by Gaussian beams. By analytically describing the lift-off area as a function of peak fluence, beam radius, and focus position, we demonstrate that the maximum lifted-off area is achieved at a substantially lower optimal fluence, namely $F_0^{\text{opt}} = e^1 F_{\text{th}}$. Closed-form expressions for the optimal beam radius, maximal lift-off area, and optimal focus position are derived and validated by numerical modeling. The theory is applied to picosecond laser lift-off of copper oxide from copper, showing excellent agreement between experimental observations and model predictions. The results reveal fundamental differences between ablation- and lift-off-dominated material removal and provide practical guidelines for maximizing process efficiency in laser-assisted delamination, selective coating removal, and surface functionalization.

## 1. Introduction

Laser–matter interaction has become an indispensable tool in advanced manufacturing, enabling precise material removal, surface structuring, and functional modification across a wide range of length scales [1–4]. In particular, ultrafast burst and short-pulse laser processing offers unique advantages due to the strong confinement of energy deposition, high spatial selectivity, and reduced thermal damage [5–7]. These properties have led to widespread adoption of laser-based techniques for microelectronics fabrication, biomedical device manufacturing, and selective surface engineering [8,9].

Among the various laser-induced material removal mechanisms [10–12], laser ablation has been studied extensively, and its efficiency limits are well understood [13]. A classical result of ablation theory shows that, for a Gaussian beam, the maximum ablated volume per pulse is achieved at a specific peak fluence that exceeds the ablation threshold by a constant exponential factor [14,15]. This concept of *efficient* *ablation* has provided a powerful framework for optimizing laser processes by balancing energy utilization against material removal yield and has been validated across different materials and pulse durations [16–19].

In contrast, laser-induced lift-off and delamination processes—where a surface layer is removed without complete volumetric ablation of the underlying substrate—remain significantly less explored from a fundamental optimization perspective [20,21]. Laser lift-off plays a critical role in the selective removal of oxides [20,21], thin films [22,23], solar modules [24], graphene [25], and functional layers [22,26], where controlled interfacial separation is desired rather than deep material excavation. In such cases, material removal is governed not only by optical absorption but also by thermomechanical stress generation, interfacial adhesion, and fracture dynamics [21,27].

Despite its technological relevance, laser lift-off is often optimized empirically, with processing parameters selected based on trial-and-error approaches. While threshold fluence values for lift-off are commonly reported [20,21], a universal criterion for maximizing the lifted-off area per pulse has not been established. Moreover, optimization strategies derived from ablation theory are frequently applied to lift-off processes without considering the fundamentally different physics involved [23], potentially leading to suboptimal energy usage and process inefficiency.

From a physical standpoint, lift-off differs from ablation in that the removed material does not necessarily undergo complete phase transition or vaporization [28]. Instead, the process is driven by rapid energy deposition within a confined volume, leading to stress accumulation and subsequent interfacial failure [21]. As a result, the relationship between laser fluence and removal efficiency is expected to deviate from that of classical ablation, particularly in the regime just above the lift-off threshold.

In this work, we address this gap by developing an analytical theory of efficient laser lift-off based on Gaussian beam optics and threshold-driven material response. By describing the lifted-off area as a function of peak fluence, beam radius, and focal position, we derive closed-form expressions for the conditions that maximize the lift-off area per laser pulse. In direct analogy to efficient ablation theory, we identify a distinct optimal fluence condition for lift-off, revealing a fundamental exponential scaling with the lift-off threshold fluence.

The theoretical framework is applied to picosecond laser lift-off of copper oxide from a copper substrate, combining analytical modelling, numerical simulations, and experimental validation. The results demonstrate that the optimal fluence for maximal lift-off area occurs at a substantially lower value than that required for efficient ablation, highlighting the fundamentally different nature of delamination-dominated material removal. The presented model provides clear and practical guidelines for optimizing laser lift-off processes and establishes efficient lift-off as a complementary concept to efficient ablation in laser–matter interaction.

## 2. Model

### 2.1. Laser lift-off area by Gaussian beam

We assume that the laser spot has a Gaussian intensity profile with a spot radius $w$ (defined at the $e^{-2}$ intensity level). Thus, the fluence distribution at a radial position $r$ on the material surface can be expressed as [29]:

$$F(r) = F_0 \exp\left(\frac{-2r^2}{w^2}\right), \tag{1}$$

The peak laser fluence of a Gaussian beam $F_0$ for a single pulse is given by [29]:

$$F_0 = \frac{2E_p}{\pi w^2}, \tag{2}$$

where $E_p$ is the energy contained in a single laser pulse. Diameter squared of laser lift-off area [14,20,30]:

$$D^2 = 2w^2 \ln\left(\frac{F_0}{F_{th}}\right), \tag{3}$$

where $F_{th}$ represents the threshold fluence of laser ablation for particular investigated material. Ablated (lift-off) area can be expressed as follows:

$$A = \frac{\pi D^2}{4}. \tag{4}$$

As the lift-off area is proportional to the square of the ablated diameter (Eq. (4)), and the ablated diameter depends on the beam radius, peak fluence, and lift-off threshold (Eq. (3)), while the peak fluence itself is a function of the beam radius and pulse energy (Eq. (2)), the listed expressions can be combined to obtain the lift-off area as a function of the processing parameters. In particular, the lift-off area can be expressed in terms of the pulse energy, Gaussian beam radius, peak on-axis fluence of the Gaussian beam, and the lift-off threshold. In the following sections, we theoretically analyze the dependence of the ablated area on these parameters. We also derive expressions for its maximum values and identify the corresponding optimal processing conditions. Because the peak laser fluence depends on the beam radius, we derive the lift-off area as a function of the beam radius and determine the optimal radius for maximizing the ablated area. In addition, the Gaussian beam propagation along the $z$-axis modifies both the radius and the corresponding peak fluence; therefore, we analyze the axial dependence of the lift-off area and identify the optimal off-focus position for achieving the maximum lift-off area.

### 2.2. Optimal laser fluence for maximal lift-off area

By inserting Eq. (3) into Eq. (4):

$$A(F_0) = \frac{E_p}{F_0} \ln\left(\frac{F_0}{F_{th}}\right). \tag{5}$$

We have derived analytical expressions for the laser lift-off (ablated) area as a function of the relevant processing parameters, including pulse energy, peak on-axis fluence of the Gaussian beam, and the lift-off threshold. By calculating first derivative of Eq. (5) by peak laser fluence and equaling it to zero we get the expression to calculate the optimal peak laser fluence for maximal laser lift-off area:

$$\frac{dA}{dF_0} = \frac{E_p}{F_0^2}\left(1 - \ln\left(\frac{F_0}{F_{th}}\right)\right) = 0. \tag{6}$$

By solving Eq. (6) we obtain optimal peak laser fluence $F_0^{opt}$ in the center of Gaussian beam for maximal laser lift-off area:

$$F_0^{opt} = e^1 F_{th} \simeq 2.72 F_{th}. \tag{7}$$

By inserting Eq. (7) into Eq. (5) we obtain maximal laser lift-off (ablated) area $A_{max}$:

$$A_{max} = \frac{E_p}{e^1 F_{th}} \simeq 0.368 \frac{E_p}{F_{th}}. \tag{8}$$

The maximal lift-off area is directly proportional to pulse energy and inversely proportional to lift-off threshold.

### 2.3. Optimal beam radius for maximal lift-off (ablated) area

By inserting Eq. (2) into Eq. (5) we replace the peak laser fluence with pulse energy and Gaussian beam radius:

$$A(w) = \frac{\pi w^2}{2} \ln\left(\frac{2E_p}{\pi w^2 F_{th}}\right). \tag{9}$$

By calculating first derivative of Eq. (9) by Gaussian beam radius and equaling it to zero we get the expression to calculate the optimal beam radius for maximal laser lift-off area:

$$\frac{dA}{dw} = \pi w \left(\ln\left(\frac{2E_p}{\pi w^2 F_{th}}\right) - 1\right) = 0. \tag{10}$$

By solving Eq. (10) we achieve optimal laser beam spot radius $w_{opt}$ for maximal laser lift-off efficiency:

$$w_{opt} = \sqrt{\frac{2E_p}{e^1 \pi F_{th}}} \simeq 0.484 \sqrt{\frac{E_p}{F_{th}}}. \tag{11}$$

By inserting Eq. (11) into Eq. (9) we obtain maximal laser lift-off (ablated) area $A_{max}$:

$$A_{max} = \frac{\pi}{2} w_{opt}^2 \simeq 1.57 w_{opt}^2. \tag{12}$$

The maximal lift-off area is proportional to the optimal beam radius squared.

### 2.4. Gaussian beam propagation

To account for the influence of focusing conditions on the laser-induced lift-off process, the propagation of a real (non-ideal) Gaussian beam along the optical axis is considered. The evolution of the beam radius $w(z)$ as a function of the longitudinal coordinate $z$ is described by [29,31]:

$$w^2(z) = w_0^2 \left(1 + \left(\frac{z - z_0}{z_R}\right)^2\right), \tag{13}$$

where $z_R$ is the (generalized) Rayleigh range:

$$z_R = \frac{\pi w_0^2}{\lambda M^2}, \tag{14}$$

where $w_0$ is the beam waist radius at the focal plane, $\lambda$ is the laser wavelength, $z_0$ is the beam waist position, and $M^2$ is the beam quality factor accounting for deviations from an ideal Gaussian beam. The corresponding peak laser fluence in the center of the Gaussian beam depends on both the pulse energy and the beam radius at a given axial position. Taking into account the displacement of the sample relative to the focal plane and combining Eq. (13) and Eq. (2), the peak fluence $F_0(z)$ can be expressed as:

$$F_0(z) = \frac{2E_p}{\pi w^2(z)}. \tag{15}$$

Eq. (13) and Eq. (15) illustrate the intrinsic coupling between beam divergence and peak fluence during longitudinal displacement of the sample. As the distance from the focal plane increases, the beam radius expands while the peak fluence decreases, directly affecting the lift-off efficiency.

## 2.5. The optimal focus position for most efficient lift-off

In real-world applications, laser photons—and therefore laser power—are costly, so it is generally desirable to utilize the maximum available power in a given process. Since the laser pulse energy is typically fixed within a relatively narrow adjustment range, the most effective way to optimize energy usage is by adjusting the laser spot size by varying the focus position.

We can obtain lift-off area dependence on longitudinal $z$ position of sample surface $A = A(z)$, by having processing parameter dependences on longitudinal $z$ position Eq. (5), Eq. (13), and Eq. (15):

$$A(z) = \frac{\pi}{2} w^2(z) \ln\left(\frac{2E_p}{\pi w^2(z) F_{th}}\right). \tag{16}$$

There are several ways to determine the optimal beam waist position that maximizes the lift-off area. One approach is to evaluate the first derivative of Eq. (16) with respect to the longitudinal sample position, set it equal to zero, and solve for the optimal off-focus offset. Alternatively, one may express the optimal beam waist as a function of the longitudinal position and subsequently determine the corresponding optimal distance from the focus that yields the maximum lift-off area. For simplicity, we adopt the second approach. Longitudinal dependence of the spot size with the optimal beam waist condition, we obtain the following expression:

$$w(z_{opt}) = w_{opt}. \tag{17}$$

By equating Eq. (13) and Eq. (11), i.e., we can rewrite Eq. (17):

$$\frac{2E_p}{e^1 \pi F_{th}} = w_0^2 \left(1 + \left(\frac{z_{opt} - z_0}{z_R}\right)^2\right). \tag{18}$$

Now we can extract optimal focus off set distance $z_{opt}$ as a solution of Eq. (18):

$$z_{opt} = z_0 \pm z_R \sqrt{\frac{2E_p}{e^1 \pi w_0^2 F_{th}} - 1}. \tag{19}$$

By inserting Eq. (19) into Eq. (16) we obtain maximal laser ablated (lift-off) area $A_{max}$:

$$A_{max} = \frac{E_p}{e^1 F_{th}} \simeq 0.368 \frac{E_p}{F_{th}}. \tag{20}$$

The derived Eq. (20) is identical to Eq. (8), confirming that the maximal lift-off area can be achieved by adjusting the peak laser fluence through variation of the sample position $z$ relative to the beam waist.

## 3. Results

### 3.1. Numerical simulation of optimal laser lift-off

The numerical modelling of copper oxide lift-off area depending on pulse energy, sample focus position, beam spot radius and the peak laser fluence in center of Gaussian beam calculated by using Eq. (5), Eq. (9), Eq. (16). Fig. 1 summarizes the calculated dependence of the oxide lift-off area $A$ on the pulse energy $E_p$, the beam waist $w$, the longitudinal sample position $z$, and the peak on-axis fluence $F_0$.

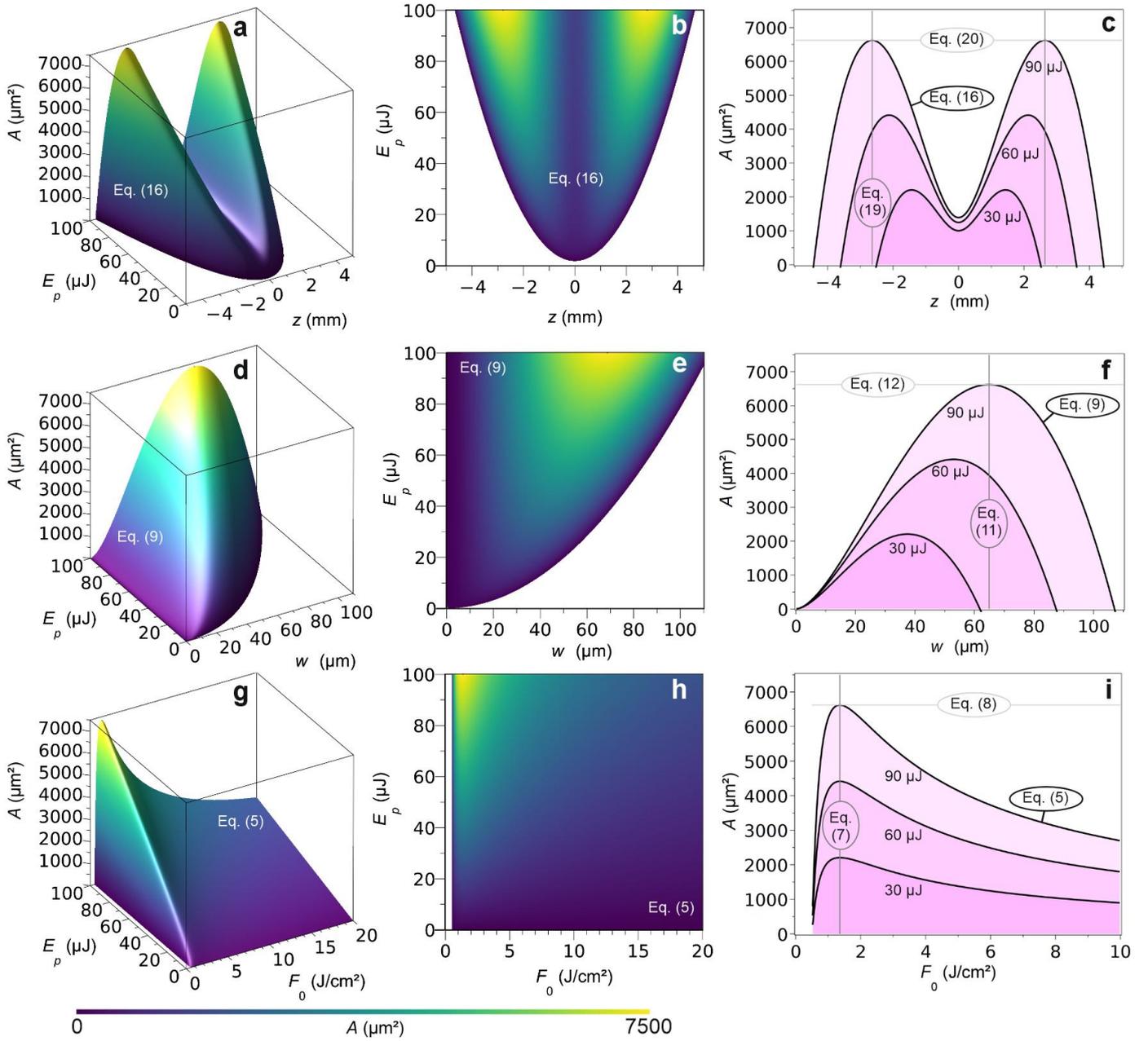

**Fig. 1** Calculated copper oxide lift-off area as a function of laser processing parameters. Laser lift-off area (color scale) dependence on the pulse energy, longitudinal sample position z, beam radius w, and peak on-axis fluence $F_0$: (**a, d, g**) 3D views and (**b, e, h**) 2D maps of modeling results. (**c, f, i**) 1D profiles of lift-off area dependence on sample focus off-set position, beam radius, and peak laser fluence at different pulse energies. The color scale at the bottom of the figure corresponds to panels (**a, b, d, e, g, h**). Model parameters used in modelling: laser wavelength $\lambda$ = 1064 nm, copper oxide lift-off threshold $F_{th}$ = 0.5 J/cm², the beam quality parameter M² = 1.06, focused beam waist radius $w_0$ = 15 µm, Rayleigh length $z_R$ = 0.63 mm, beam waist position $z_0$ = 0 mm.

The model uses a Gaussian beam with central wavelength $\lambda$ = 1064 nm, copper oxide lift-off threshold fluence $F_{th}$ = 0.5 J/cm², and beam quality parameter M² = 1.06, focused beam radius on the sample $w_0$ = 15 µm, Rayleigh length $z_R$ = 0.63 mm, beam waist position $z_0$ = 0 mm.

Tho computational results of lift-of area depending on pulse energy and peak laser fluence calculated by using Eq. (16) is provided in Fig. 1(a–c). It shows that for a fixed pulse energy, the lift-off area strongly depends on the longitudinal $z$ position relative to Gaussian beam waist position. Maximum $A$ occurs symmetrically around the best sample position $z_{\text{opt}}$, defined by Eq. (19) due to the increased local fluence and reduced spot size, and increases monotonically with pulse energy. And the maximal lift-off area can be evaluated by using Eq. (20).

The calculated lift-of area depending on pulse energy and beam radius calculated by using Eq. (9) is shown in Fig. 1(d–f). It demonstrates that increasing the beam waist increases the lift-off area up to an optimum radius defined by Eq. (11), above which $A$ decreases as the fluence becomes sub-threshold across most of the profile. The maximal lift-off area can be evaluated by Eq. (12).

The computed lift-of area depending on pulse energy and peak laser fluence calculated by using Eq. (5) is depicted in Fig. 1(g–i). Instead parameterize the results in terms of the peak fluence $F_0$, revealing that for a given pulse energy, there exists a characteristic fluence, defined by Eq. (7) that maximizes the lift-off area, witch can be defined by Eq. (8). Below threshold the lift-off vanishes, while at high $F_0$ the effective interaction area decreases due to the confining spot geometry.

Together, these simulations provide a quantitative map of the processing parameter space and predict the existence of optimal combinations of $E_{\text{p}}$, $w$, and $z$ that maximizes the lift-off area for a given material system.

### 3.2. Laser lift-off and beam characterization experiment

The experiments were performed on copper (Cu) samples covered with a native cuprous oxide (Cu$_2$O) layer. The copper substrate (CW004A, Ekstremalė) had a mirror-like surface finish with an average roughness of $R_{\text{a}} < 0.1$ µm, measured using a stylus profiler (Dektak 150, Veeco). Laser processing was carried out using a picosecond laser system (Atlantic, Ekspla) operating at a wavelength of $\lambda = 1064$ nm with a pulse duration of $\tau_{\text{FWHM}} = 10$ ps. The laser delivered pulse energies up to $E_{\text{p}} = 142$ µJ at a repetition rate of $f_{\text{rep}} = 100$ kHz. The laser manufacturer reported Gaussian beam quality factors (M²) of 1.062 and 1.043 in the transverse $x$ and $y$ directions, respectively. Single-pulse irradiation conditions were used in all experiments. The morphology of the laser-processed oxide/metal interface was characterized by scanning electron microscopy (SEM) (JSM-6490LV, JEOL) and optical profilometry (S neox, Sensofar), as shown in Fig. 2(a–c). The experimental procedure, including the $z$-scan methodology used for lift-off investigations, follows the approach described in our previous work [20].

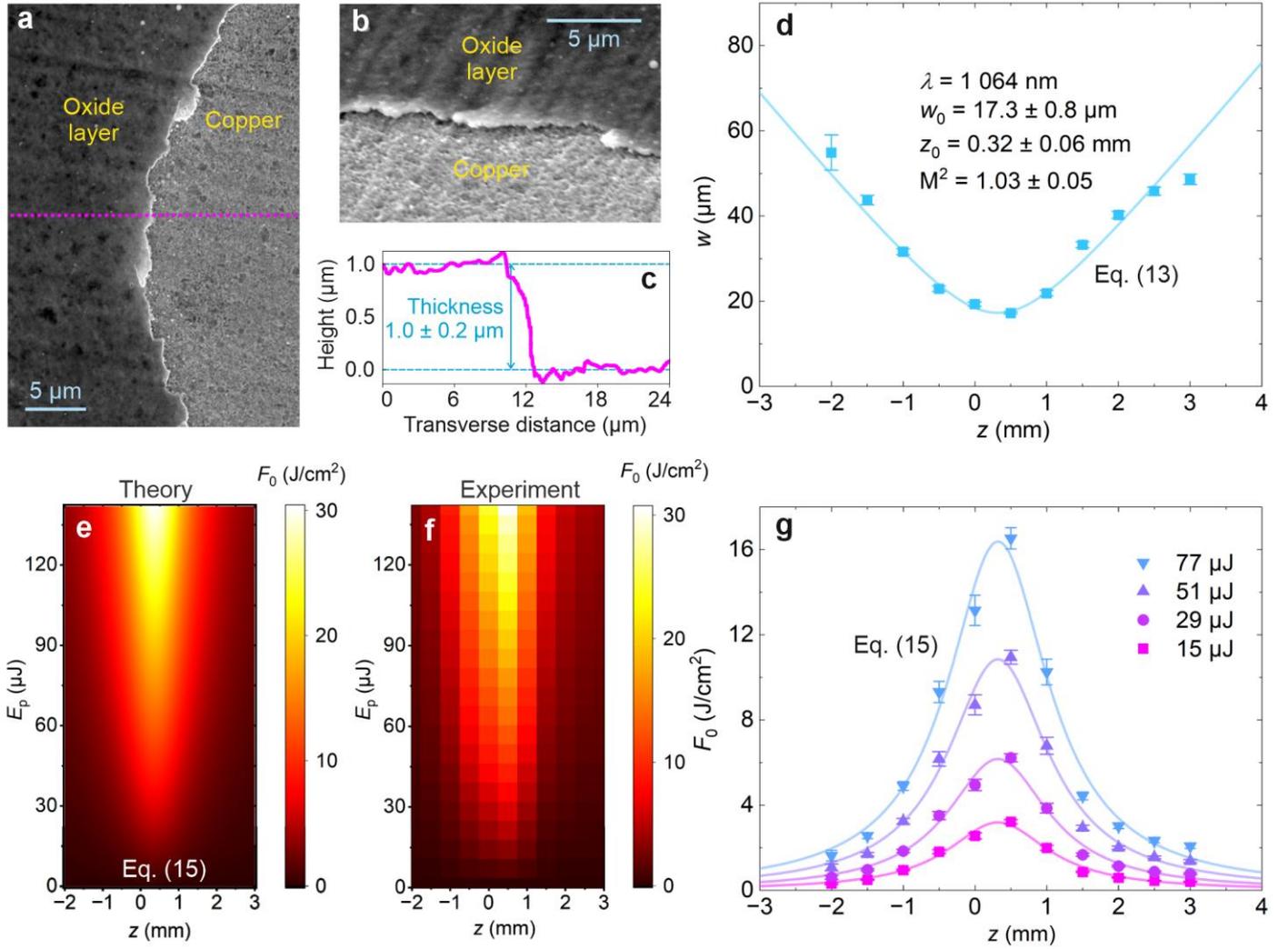

**Fig. 2** (**a–c**) Example of laser lift-off of an oxide layer from copper: (**a**) SEM micrograph showing the interface between the oxide layer and the copper substrate; (**b**) higher-magnification view acquired at a tilt angle of 45°; (**c**) height profile extracted along the dotted line in (**a**), showing the step between the oxide layer and the copper surface. The oxide layer thickness determined from the profile is approximately $1.0 \pm 0.2$ μm. (**d–g**) Gaussian beam propagation characteristics and resulting fluence distribution: (**d**) experimentally measured Gaussian beam radius as a function of sample position; solid dots with error bars – experimental data, solid line – fit by Eq. (13); (**e**) theoretical map of the peak on-axis fluence $F_0$ as a function of pulse energy $E_p$ and sample position $z$, calculated using Eq. (15) with Gaussian beam parameters obtained from the fit in panel (**d**); (**f**) experimentally reconstructed map of the peak on-axis fluence $F_0$ as a function of pulse energy $E_p$ and sample position $z$; the fluence values were calculated using Eq. (2) together with the beam radius obtained in panel (**d**); (**g**) profiles of the peak on-axis fluence $F_0$ as a function of sample position; solid dots with error bars – experimental values calculated using Eq. (2), solid lines – fit by Eq. (15).

The beam radius on the copper sample was evaluated using the classical Liu method [30]. The experimentally measured Gaussian beam radius as a function of sample position is shown in Fig. 2(d). The dependence of the beam radius on the vertical sample position was fitted using Eq. (13), which describes the propagation of a non-ideal Gaussian beam. The fit shows excellent agreement with the experimental data, confirming the validity of the Gaussian beam approximation under the applied conditions. From the fitting procedure, the beam waist radius $w_0$, the beam quality factor $M^2$, and the focal position $z_0$ were determined. The beam quality factor was found to be $M^2 = 1.03 \pm 0.05$, indicating near-diffraction-limited beam propagation. This value is consistent with values provided by laser manufacturer and obtained in our previous measurements [29]. The corresponding beam waist radius and focal position were determined as $w_0 = 17.3 \pm 0.8$ μm and $z_0 = 0.32 \pm 0.05$ mm, respectively. The retrieved parameters were subsequently used to calculate the axial evolution of the peak laser fluence using Eq. (2), as well as its theoretical propagation according to Eq. (15), providing a consistent description of both beam divergence and fluence redistribution along the propagation axis. This approach enables a direct correlation between the experimental lift-off behavior and the theoretically predicted optimal processing conditions.

Fig. 2 (e–f) summarizes the Gaussian beam propagation characteristics used in the model defined by Eq. (13) and Eq. (15), illustrating the variation of the beam radius with the sample position $z$ and the corresponding evolution of the peak on-axis fluence $F_0$. Fig. 2(e) presents the theoretical fluence distribution, while Fig. 2(f) shows the experimentally reconstructed map derived from the measured beam parameters. The good agreement between the theoretical (Fig. 2(e)) and experimental (Fig. 2(f)) results confirms the validity of the Gaussian beam propagation model used for subsequent lift-off analysis. Fig. 2(g) presents the profiles of the peak on-axis fluence $F_0$ as a function of the sample position $z$, comparing experimentally evaluated values with the theoretical fit obtained from Eq. (15). The experimental fluence values were calculated using Eq. (2) based on the measured beam radii shown in Fig. 2(d). The profiles exhibit very good agreement with the theoretical prediction, confirming the consistency of the fluence reconstruction and the validity of the Gaussian beam propagation model.

### 3.3. Optimal laser lift-off: experiment versus model

Fig. 1 presents a comprehensive comparison between experimental results and theoretical predictions of the copper oxide lift-off area $A$ as a function of key laser processing parameters, including pulse energy $E_\mathrm{p}$, sample position relative to the focal plane $z$, beam radius $w$, and peak on-axis fluence $F_0$.

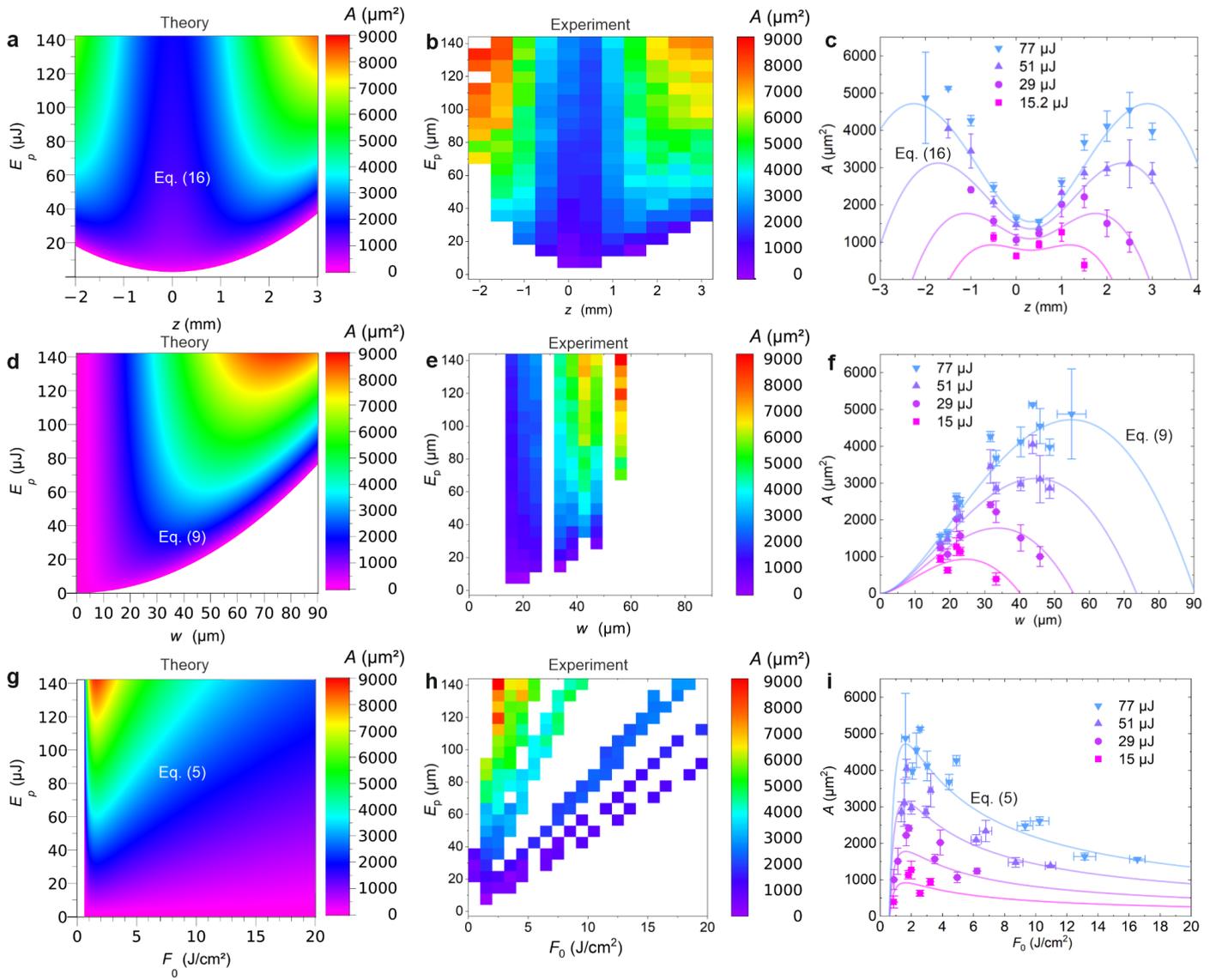

**Fig. 3** (**a–i**) Oxide lift-off area $A$ as a function of laser processing parameters: experiment versus model. (**a, d, g**) Theoretical maps of the lift-off area calculated using the Gaussian beam model, showing the dependence of $A$ on pulse energy $E_\mathrm{p}$ and: (**a**) sample position $z$ (Eq. (16)); (**d**) beam radius $w$ (Eq. (9)); (**g**) peak on-axis fluence $F_0$ (Eq. (5)). (**b, e, h**) Experimentally measured lift-off area under corresponding irradiation conditions. (**c, f, i**) One-dimensional profiles of the lift-off area as a function of key parameters: (**c**) sample position $z$; (**f**) beam radius $w$; (**i**) peak on-axis fluence $F_0$. Symbols represent experimental data for different pulse energies, while solid lines show the model predictions based on Eq. (16), Eq. (9), and Eq. (5), respectively. The color scale represents the lift-off area $A$. The experimental trends follow the theoretical predictions, with maximum lift-off areas observed at off-focus positions, intermediate beam radii, and fluence values close to the theoretically predicted optimum. Laser parameters used in experiments and modeling: pulse duration $\tau_\mathrm{FWHM} = 10$ ps, wavelength $\lambda = 1064$ nm, pulse repetition rate $f_\mathrm{rep} = 100$ kHz (pulse separation $t_\mathrm{rep} = 10$ μs), and beam quality factor $\mathrm{M}^2 = 1.03$, oxide lift-off threshold fluence $F_\mathrm{th} = 0.6$ J/cm$^2$, focused beam waist radius $w_0 = 17.3$ μm, Rayleigh length $z_\mathrm{R} = 0.86$ mm, beam waist position $z_0 = 0.32$ mm.

The experimentally determined beam radii (Fig. 2(d)) and corresponding peak fluence values (Fig. 2(g)) were used as input parameters for the efficient lift-off analysis. The experimental value of the copper oxide lift-off from copper threshold fluence $F_{th} = 0.6$ J/cm$^2$, obtained in our previous work [20], together with the beam parameters determined in Fig. 2(d) ($w_0 = 17.3$ μm, $\lambda = 1064$ nm, $M^2 = 1.03$, $z_R = 0.86$ mm, $z_0 = 0.32$ mm), were used as input parameters in the theoretical modeling presented in Fig. 3(a, d, g) and Fig. 3(c, f, i).

Fig. 3(a–c) shows the dependence of the lifted-off area $A$ on the sample position $z$ relative to the laser focal plane for different pulse energies. For all investigated pulse energies, the lift-off area exhibits pronounced maxima at off-focus positions, while a minimum is observed near the focal plane ($z = z_0 \approx 0.32$ mm). This behavior confirms that optimal lift-off efficiency is achieved under conditions where the beam radius and peak fluence jointly satisfy the theoretical optimum for maximal lift-off area derived above.

The two-dimensional maps in Fig. 3(a, b) provide a direct comparison between the theoretically predicted and experimentally measured lift-off area as a function of pulse energy $E_p$ and sample position $z$. The theoretical map (Fig. 3(a)) reveals a characteristic double-lobed distribution of the lift-off area, with maxima located symmetrically on both sides of the focal plane and increasing with pulse energy. The experimentally reconstructed map (Fig. 3(b)) reproduces the same overall trends, including the position of the maxima, the central minimum near focus, and the expansion of the high-efficiency region with increasing pulse energy. Despite minor deviations due to experimental uncertainties and discretization of the measurement grid, the agreement between the theoretical and experimental maps is very good. This confirms that the model captures the essential physics governing the lift-off process and further supports the validity of the analytical description of efficient lift-off under varying focusing conditions.

With increasing pulse energy, the maximum lift-off area increases systematically, and the position of the optimum shifts slightly away from the focal plane. At the highest investigated pulse energy of 142 μJ, the maximal lift-off area reaches ~ 9000 μm$^2$, which is close to the theoretical value of ~ 8700 μm$^2$ calculated using Eq. (20), whereas at lower energies the corresponding maxima decrease approximately proportionally, indicating that the process remains in a threshold-dominated regime. For all energies, the experimentally observed minimum near focus indicates that the beam spot radius becomes too small, resulting in fluence values significantly exceeding the optimal condition $F_0^{opt} = e^1 F_{th}$ and thus inefficient energy utilization for lift-off.

The solid curves in Fig. 3(c) represent the theoretical predictions obtained by combining the Gaussian beam propagation model with the analytical expression for efficient lift-off given by Eq. (16). The model reproduces the key features of the experimental data, including the symmetric maxima on both sides of the focal plane, the depth of the central minimum, and the dependence of the maximum area on pulse energy. The very good agreement between experiment and theory confirms that the interplay between beam divergence and peak fluence governs the lift-off efficiency and validates the analytical optimum conditions established in the previous section. Overall, the results presented in Fig. 3(c) provide direct experimental evidence that the efficient laser lift-off regime is achieved not at the focal plane, but at positions where the beam radius increases sufficiently to reduce the peak fluence to the theoretically predicted optimal value.

Fig. 3(d–f) presents the lifted-off area $A$ as a function of the laser beam spot radius $w$ for several pulse energies. In this representation, the geometric focusing dependence is decoupled from the axial position by converting the measured $z$ values into the corresponding spot radii using the Gaussian beam propagation model (Fig. 2(d)). For each pulse energy, the lift-off area exhibits a well-defined maximum at an intermediate spot size, decreasing toward both smaller and larger $w$ values. This behavior reflects the existence of an optimal balance between peak fluence and irradiated area, as predicted by the analytical efficient lift-off theory by Eq. (9).

The two-dimensional maps in Fig. 3(d, e) provide a direct comparison between the theoretically predicted and experimentally measured lift-off area as a function of pulse energy $E_\text{p}$ and beam radius $w$. The theoretical map (Fig. 3(d)) shows a well-defined optimal region at intermediate spot sizes, which shifts toward larger $w$ values with increasing pulse energy. The experimentally obtained map (Fig. 3(e)) reproduces these trends, including the location of the maxima and the overall shape of the high-efficiency region. The good agreement between theory and experiment confirms that the beam-radius-based representation accurately captures the conditions for efficient lift-off and validates the analytical model defined by Eq. (9).

At small spot radii, the peak fluence significantly exceeds the optimal value $F_0^\text{opt} = e^1 F_\text{th} \cong 1.6 \text{ J/cm}^2$, evaluated by Eq. (7), resulting in inefficient utilization of the deposited energy and a reduced lift-off area. Conversely, at larger spot radii, the fluence decreases below the threshold, again leading to reduced lift-off. The optimal regime is reached when the spot size and fluence simultaneously satisfy the theoretical efficiency condition, yielding maximal area removal. The optimal spot radius $w_\text{opt}$ increases with pulse energy, consistent with the theoretical expectation that higher pulse energies require larger beam radii to maintain the fluence at the optimal level.

The solid curves in Fig. 3(f) represent the theoretical predictions of the lift-off model evaluated as a function of spot radius for each pulse energy, based on Eq. (9). The model reproduces both the shape and magnitude of the experimental trends, including the position of the maxima and the scaling of the lifted-off area with pulse energy. This agreement confirms that the spot-radius-based representation provides a more direct visualization of the efficient lift-off condition than the axial coordinate, as it isolates the parameter that explicitly enters the analytical expression for the optimal area.

Overall, these results demonstrate that the efficient lift-off regime is achieved at a well-defined spot radius determined by the pulse energy and beam propagation parameters. Operating under these optimal conditions enables a substantial increase in lift-off efficiency compared to conventional processing at the focal plane.

Fig. 3(g–i) shows the lifted-off area $A$ as a function of the peak on-axis fluence $F_0$ for different pulse energies. The fluence values were obtained by converting the axial sample position $z$ into the corresponding $F_0(z)$ using the Gaussian beam propagation model (Fig. 2(e–g)). This fluence-based representation enables a direct comparison with the analytical efficient lift-off theory, which predicts a well-defined optimum at $F_0^{\text{opt}} = e^1\, F_{\text{th}}$.

For all investigated pulse energies, the lift-off area exhibits a pronounced maximum at intermediate fluence values, followed by a monotonic decrease at higher fluence. This behavior represents the key experimental signature of the efficient lift-off regime. At low fluence, the illuminated area only marginally exceeds the threshold, leading to incomplete delamination. In contrast, at high fluence, the illuminated area becomes too small due to strong beam focusing, resulting in inefficient energy utilization.

In contrast to efficient ablation, where the optimal fluence occurs significantly above the threshold ($F_0^{\text{opt}} = e^2 F_{\text{th}}$), the lift-off optimum is located closer to the threshold fluence due to the dominant role of interfacial stress and fracture processes.

The solid curves in Fig. 3(i) represent the analytical model evaluated as a function of $F_0$ for each pulse energy, based on Eq. (5). The model accurately reproduces both the position of the maxima and the characteristic asymmetry of the experimental curves, confirming that the peak fluence is the primary control parameter governing lift-off efficiency. The good agreement between experiment and theory demonstrates that efficient lift-off is achieved by adjusting the beam propagation conditions such that the fluence reaches $F_0^{\text{opt}} = e^1\, F_{\text{th}}$ at off-focus positions rather than at the focal plane.

Overall, this fluence-based representation highlights the practical implications of efficient lift-off theory: maximum lift-off efficiency occurs not at the highest achievable fluence but rather at a reduced fluence value, determined by a balance between illuminated area and detachment threshold. This finding provides a clear guideline for process optimization and explains why conventional on-focus processing approaches can be energetically suboptimal for lift-off-dominated material removal.

**Conclusions**

In this work, we developed a theoretical framework for efficient laser-induced lift-off of surface layers, extending the classical concept of efficient laser ablation to delamination-dominated material removal. By analytically describing the lift-off area produced by a Gaussian laser beam, we derived closed-form expressions linking the ablated (lift-off) area to peak fluence, beam radius, and focus position. A central result of this study is the identification of a new optimal fluence condition for maximal lift-off efficiency, showing that the maximum lifted-off area is achieved at a peak fluence of $F_0^{\text{opt}} = e^1 F_{\text{th}}$, which is fundamentally lower than the well-known $F_0^{\text{opt}} = e^2 F_{\text{th}}$ condition for maximum ablated volume in efficient ablation theory. This difference highlights the distinct physical nature of lift-off processes, where interfacial failure and stress-driven delamination dominate over bulk material removal.

The analytical model further provides explicit expressions for the optimal beam radius, laser fluence, and focus position that maximize the achievable lift-off area, accounting for non-ideal Gaussian beam propagation. Numerical simulations reveal well-defined maxima in the lift-off area as a function of laser fluence, beam radius, and focusing conditions, providing clear guidelines for process optimization. Experimental validation using picosecond laser lift-off of copper oxide from copper demonstrates very good agreement with the theoretical predictions, confirming both the existence of the predicted optimal fluence and the robustness of the model across a broad parameter space. The results show that operating at the derived optimal conditions enables a significant increase in lift-off efficiency while avoiding unnecessary energy deposition.

Overall, the presented theory establishes efficient laser lift-off as a distinct optimization approach, complementary to efficient ablation theory. The findings provide a universal and practical framework for the optimization of laser-assisted delamination, selective oxide and coating removal, and surface functionalization processes, with direct relevance to microelectronics, biomedical device fabrication, and advanced manufacturing applications.

**Author contributions**

A.Ž., P.G., and M.G. conceived the original research idea. P.G. secured the funding, supervised the project, and provided access to the state-of-the-art laser microfabrication system. A.Ž. designed and performed the laboratory experiments. M.G. analyzed the experimental data, developed the theoretical model, performed the numerical calculations, compared the experimental and numerical results, and wrote the manuscript. All authors discussed the results and contributed to the final manuscript.

**Acknowledgements**

This project has received funding from the Research Council of Lithuania (LMTLT), agreement No. S-IMPRESSU-24-1.

**Funding**

Research Council of Lithuania (LMTLT) (S-IMPRESSU-24-1).